\documentclass[prb,preprint,showpacs,preprintnumbers,amsmath,amssymb]{revtex4}

\usepackage[dvipdfmx]{graphicx}
\usepackage{bm}

\newcommand{\be}{\begin{equation}}
\newcommand{\ee}{\end{equation}}
\newcommand{\eq}[1]{Eq.~(\ref{#1})}
\newcommand{\fig}[1]{Fig.~\ref{#1}}
\def\bea{\begin{eqnarray}}
\def\eea{\end{eqnarray}}

\def\vk{{\bf k}}
\def\vQ{{\bf Q}}
\def\vr{{\bf r}}

\begin{document}

\title{Anomaly of longitudinal spin susceptibility at superconducting instability inside a magnetic phase} 

\author{Hiroyuki Yamase$^{1,2}$ and Muhammad Zafur$^{2,1}$}
\affiliation{
{$^{1}$}International Center of Materials Nanoarchitectonics, 
National Institute for Materials Science, Tsukuba 305-0047, Japan \\ 
{$^{2}$}Department of Condensed Matter Physics, Graduate School of Science, 
Hokkaido University, Sapporo 060-0810, Japan 
}

\date{March 2, 2021}

\begin{abstract}
We study the longitudinal spin susceptibility inside a magnetically ordered phase, which exhibits 
a superconducting instability leading to a coexistence of  the two ordered phases. 
Inside the magnetic phase,  the superconducting gap acquires a linear term in a magnetic field 
applied along the direction of the magnetic moment. 
We find that such a linear term generates a jump of the longitudinal spin susceptibility 
when the superconducting instability occurs via a continuous phase transition. 
This anomaly at the superconducting instability 
is a thermodynamic signature of the microscopic coexistence of superconductivity and magnetism,  
and can be a general feature associated with the breaking of spin rotational symmetry inside the magnetic phase. 
\end{abstract}


\maketitle
\section{introduction}
Magnetic systems are often envisaged as insulators characterized by a local moment   
at magnetic ion sites. However, metallic systems can also be magnetized and a spin-density-wave 
is known as a central concept to describe the itinerant magnetism. 
In contrast to the insulating systems, both spin and charge degrees of freedom become 
active and they even couple to each other. 
A famous example is chromium \cite{fawcett88} where the incommensurate spin-density-wave 
is accompanied by the charge-density-wave whose modulation vector is twice as large as 
that of the spin-density-wave.  
The same coupled state is also known in La-based high-$T_c$ cuprates and frequently refereed to as 
spin-charge stripe order \cite{tranquada95}. 

Itinerant magnetic systems can also exhibit superconducting instabilities,  
leading to a state where the two phases coexist. 
This possibility is discussed in various materials such as cuprates \cite{niedermayer98,kimura99,haug10}, 
iron pnictides \cite{stewart11}, and heavy electron compounds \cite{pfleiderer09}. 
It is frequently controversial whether the magnetism indeed coexists with superconductivity microscopically 
or they are simply phase-separated with a possible overlap around the boundary.  
Here by "microscopically" we mean that the same electrons play a dual role leading to both 
superconductivity and magnetism. 

In multilayer cuprate superconductors, early NMR measurements reported 
a microscopic coexistence of superconductivity and magnetism inside a CuO$_2$ plane \cite{mukuda12}. 
Very recently angle-resolved photoemission spectroscopy 
reported clear evidence 
of the microscopic coexistence by observing the superconducting gap along the Fermi-surface pocket 
reconstructed by antiferromagnetic order \cite{kunisada20}. 

What can be a physical quantity which characterizes the coexistence of superconductivity and magnetism 
from a thermodynamic point of view? 
In experiments, the onset of superconductivity is monitored by the zero resistivity, Meissner effect, 
and the specific heat jump even inside a magnetic phase. However, this does not necessarily indicate 
a microscopic coexistence of superconductivity and magnetism because possible phase separation 
cannot be excluded. This is the major reason why the possible coexistence of superconductivity and magnetism 
frequently becomes a controversial issue. 

Aiming for a fundamental insight into the coexistence, we study the longitudinal spin susceptibility, which 
is a thermodynamic quantity to characterize the magnetic property of a material. 
A way to compute the spin susceptibility is already well established in the normal phase where 
no magnetic order is present \cite{mahan}. However, it is not necessarily well recognized how one should compute 
the spin susceptibility in a magnetically ordered phase. A standard procedure is to compute a bubble 
diagram, including Umklapp processes when the translational symmetry is broken by a magnetic order, 
in terms of Green's functions of quasiparticles defined inside the magnetic phase. 
The effect of interactions between quasiparticles is then considered frequently in 
the random phase approximation (RPA). 
This procedure may yield correct results for the transverse spin susceptibility, but not necessarily for the longitudinal 
spin susceptibility. The point lies in a fact that because of breaking of spin rotational symmetry inside 
the magnetic phase  the electron density or the chemical potential is no longer a quadratic function 
of a magnetic field and acquires 
a linear term in the field when computing the longitudinal spin susceptibility \cite{kuboki17}. 
Since the spin susceptibility is a linear response quantity to a magnetic field, such an emergent linear term 
should be taken into account and it actually plays a crucial role \cite{kuboki17}. 
However, many studies \cite{sokoloff69,schrieffer89,kampf94,knolle10,rowe12} missed the contribution of 
additional linear terms in a field other than the magnetism. 
It is only a few studies \cite{chubukov92,hjlee12,kuboki17,delre21} that properly take it into account. 

The electron density and the chemical potential are different thermodynamic variables. As a result, 
the longitudinal spin susceptibility for a fixed density ($\chi_n$) can become different 
from that for a fixed chemical potential ($\chi_{\mu}$). 
It was pointed out in Ref.~\onlinecite{kuboki17} that $\chi_n$ and $\chi_{\mu}$ are connected with each other via 
a thermodynamic relation, 
\be
\chi_{n} = \chi_{\mu} + \left.\frac{\partial n}{\partial h} \right|_{\mu}  \left.\frac{\partial \mu}{\partial h} \right|_{n}\,,
\label{thermodynamics} 
\ee
where $n$ and $\mu$ are the electron density and the chemical potential, respectively. 
Here $h$ is an infinitesimally small magnetic field. Note that it does not necessarily imply a uniform field. 
More generally, $h$ is defined as $h_i = h {\rm e}^{i \vQ\cdot \vr_i}$ and $\vr_i$ runs over a lattice. 
The momentum $\vQ$ (including the case of $\vQ={\bf 0}$) describes the modulation vector of the magnetic order,  
i.e., ${m}_i = m {\rm e}^{i \vQ\cdot \vr_i}$. 
The direction of $h$ is chosen along the axis of easy magnetization. 
In this sense, the longitudinal susceptibility is defined as 
$\chi_{n (\mu)} = \lim_{h \rightarrow 0} \left. \frac{\partial m}{\partial h} \right|_{n(\mu)}$. 
Equation~(\ref{thermodynamics}) is easily obtained in thermodynamics, 
but is not recognized well. 
It immediately leads to the following. 
i) $\chi_{\mu} \geq \chi_{n}$ since 
$\left.\frac{\partial n}{\partial h} \right|_{\mu} \left.\frac{\partial \mu}{\partial h} \right|_{n}=
- \left(\left.\frac{\partial n}{\partial h} \right|_{\mu}\right)^2 \left.\frac{\partial \mu}{\partial n} \right|_{h}$ 
and the stability condition indicates that $\left.\frac{\partial \mu}{\partial n} \right|_{h}$ should be positive-semidefinite. 
ii) $\chi_{\mu} = \chi_{n}$ in an insulating system because $\frac{\partial n}{\partial h}$ vanishes due to the presence of a charge gap.  
iii) $\chi_{\mu}$ and $\chi_n$ can be different when $n$ and $\mu$ acquire a linear term in $h$. 
It is this case when a special care is required. The linear term emerges under the two conditions \cite{kuboki17}: 
(a) the system is in a magnetically ordered phase and breaks spin rotational symmetry and 
(b) $h$ is applied along the direction of the magnetic moment, which is the case in the longitudinal spin susceptibility. 

In this paper, we perform explicit calculations of $\chi_{n}$ and $\chi_{\mu}$ by focusing on a magnetic phase 
characterized by $\vQ=(\pi,\pi)$, namely the antiferromagnetic phase.  
Because of the additional linear terms in $h$ in \eq{thermodynamics}, calculations of $\chi_{\mu}$ become involved 
even in the RPA in a standard diagramatic technique \cite{chubukov92,hjlee12,kuboki17,delre21} and those of $\chi_n$ 
are more elusive. Since the RPA is equivalent to a mean-field approximation, 
we employ  a mean-field theory, which allows transparent calculations 
for both $\chi_{\mu}$ and $\chi_n$ not only in a magnetic phase but also in a coexistence phase of superconductivity 
and magnetism. We find that the longitudinal spin susceptibility exhibits a jump at the 
superconducting instability via a continuous phase transition inside the magnetic phase. 
It can be argued that this jump is a general feature independent of 
approximations and models, and a manifestation of microscopic coexistence of superconductivity and magnetism.

\section{Model and Formalism}
The coexistence of superconductivity and magnetism is obtained in various two-dimensional models such as 
$t$-$J$ (Refs.~\onlinecite{giamarchi91,inaba96,himeda99,sushkov04,shih04,yamase04a}) and Hubbard  \cite{inui88,giamarchi91,lichtenstein00,senechal05,aichhorn06,capone06,reiss07,kancharla08,jwang14,bxzheng16,yamase16} models, 
mainly motivated by the cuprate physics. 
Our finding of a jump in the longitudinal spin susceptibility may not depend on details of 
approximations and models. We therefore study a minimal model to describe the coexistence of superconductivity 
and magnetism, where itinerant electrons with a dispersion $\xi_{\vk}$ interact with each other via 
a singlet pairing interaction with strength $V_s (<0)$ and an antiferromagnetic interaction with $V_m (>0)$: 
\be
\mathcal{H}=\sum_{{\bf k} \sigma} \xi_{{\bf k}} c_{{\bf k} \sigma}^{\dagger}  c_{{\bf k} \sigma} 
+ V_s \sum_{i \tau} \hat{\Delta}_{i \tau}^{\dagger}  \hat{\Delta}_{i \tau}  
+ V_m \sum_{i \tau} \hat{m}_{i}  \hat{m}_{i+\tau} 
- \sum_{i} h_{i} \hat{m}_{i} \,.
\label{hamiltonian} 
\ee
Here $c_{\vk \sigma}^{\dagger}$ and $c_{\vk \sigma}$ are the creation and annihilation operators for electrons 
with momentum $\vk$ and spin orientation $\sigma$, respectively; $i$ runs over a square lattice and $\tau$ refers to 
the nearest-neighbor direction, i.e., $\tau=x$ and $y$; the singlet paring  and magnetization operators are defined as 
$\hat{\Delta}_{i \tau}=   c_{i \uparrow}c_{i+\tau \downarrow} - c_{i \downarrow}c_{i+\tau \uparrow}$ and 
$\hat{m}_{i} = \frac{1}{2} (c_{i \uparrow}^{\dagger} c_{i \uparrow}  -  c_{i \downarrow}^{\dagger} c_{i \downarrow})$, respectively. 
Having in mind a low-energy effective interaction, we introduce the magnetic interaction between the nearest-neighbor sites. 
While one might favor a Hubbard-like onsite interaction to describe the magnetism, 
our conclusions do not depend on such a detail. 
For a later convenience, we also introduce an infinitesimally small magnetic field 
$h_i$, which is applied along the $z$ direction and couples to the magnetic moment $\hat{m}_{i}$. 

We decouple the interaction terms by introducing mean fields: 
$\Delta_{\tau} = \langle \hat{\Delta}_{i \tau} \rangle$ and 
$m=\langle \hat{m}_i \rangle {\rm e}^{{\rm i}{\bf Q}\cdot{\bf r}_{i}}$ with $\vQ=(\pi,\pi)$ describing 
the N\'{e}el state. Those mean fields are assumed to be  uniform, not to depend on sites $i$. 
We then take the field as $h_i = h {\rm e}^{i {\bf Q} \cdot {\bf r}_i}$. 
The resulting mean-field Hamiltonian is given by 
\be
\mathcal{H}_{\rm MF} = \sideset{}{'}\sum_{{\bf k}} \Psi_{{\bf k}}^{\dagger} M_{{\bf k}} \Psi_{{\bf k}} \,,
\ee
where the summation over momentum $\vk$ is restricted to the magnetic Brillouin zone 
$|k_x| + |k_y| \leq \pi$ as indicated by prime, 
$\Psi_{{\bf k}}^{\dagger}=\left(c_{{\bf k}\,\uparrow}^{\dagger}\;\; 
c_{-{\bf k}\,\downarrow} \;\; c_{{\bf k}+{\bf Q}\,\uparrow}^{\dagger}\;\; c_{-{\bf k}+{\bf Q}\,\downarrow}\right)$, 
and  
\be
M_{{\bf k}} = \left(
 \begin{array}{cccc}
   \xi_{{\bf k}} & -\Delta_{{\bf k}}& -\overline{m} & 0 \\
-\Delta_{{\bf k}}^{*} & -\xi_{{\bf k}} & 0  & -\overline{m} \\
-\overline{m} & 0 & \xi_{{\bf k}+{\bf Q}} & -\Delta_{{\bf k}+{\bf Q}}\\
0 & -\overline{m}  &  -\Delta_{{\bf k}+{\bf Q}}^{*}& -\xi_{{\bf k}+{\bf Q}} 
\end{array}
\right) \,.
\ee
Here $\xi_{{\bf k}}=-2\left[ t (\cos k_{x}+\cos k_{y}) 
+ 2t' \cos k_{x} \cos k_{y}+
t''(\cos 2k_{x}+\cos 2k_{y})\right]  -\mu$ 
with $\mu$ being the chemical potential, $\overline{m} = 2  m V_{m} + \frac{h}{2}$, and 
$\Delta_{{\bf k}}=2 V_s \Delta_{0} \left(\cos k_{x}-\cos k_{y}\right)$.  
Since the $d$-wave superconductivity is stable in 
a parameter region we are interested in, we already put $\Delta_0=\Delta_x = -\Delta_y$ for simplicity. 
Assuming the order parameters are real, we obtain the following self-consistency equations: 
\bea
&&n=1-\frac{1}{N}\sideset{}{'}\sum_{{\bf k}}
\left(\frac{\eta_{{\bf k}}^{+}}{\lambda_{{\bf k}}^{+}}
\tanh \frac{\lambda_{{\bf k}}^{+}}{2T}+
\frac{\eta_{{\bf k}}^{-}}{\lambda_{{\bf k}}^{-}}
\tanh \frac{\lambda_{{\bf k}}^{-}}{2T} \right) \,, 
\label{self-n}\\
&&\Delta_{0}=-\frac{1}{2N}\sideset{}{'}\sum_{{\bf k}}
d_{\vk} \left(\frac{\Delta_{{\bf k}}}{\lambda_{{\bf k}}^{+}}
\tanh \frac{\lambda_{{\bf k}}^{+}}{2T}+
\frac{\Delta_{{\bf k}}}{\lambda_{{\bf k}}^{-}}
\tanh \frac{\lambda_{{\bf k}}^{-}}{2T} \right) \,,
\label{self-D} \\
&&m=\frac{1}{2N}\sideset{}{'}\sum_{{\bf k}}
\frac{\overline{m}}{D_{{\bf k}}}
\left(\frac{\eta_{{\bf k}}^{+}}{\lambda_{{\bf k}}^{+}}
\tanh \frac{\lambda_{{\bf k}}^{+}}{2T}-
\frac{\eta_{{\bf k}}^{-}}{\lambda_{{\bf k}}^{-}}
\tanh \frac{\lambda_{{\bf k}}^{-}}{2T} \right) \,,
\label{self-m}
\eea
where $n$ is the electron density per lattice site, 
$\lambda_{{\bf k}}^{\pm}=\sqrt{\eta^{\pm \; 2}_{{\bf k}}+\Delta_{{\bf k}}^{2}}$, 
$\eta_{{\bf k}}^{\pm}=\xi_{{\bf k}}^{+}\pm D_{{\bf k}}$, 
$D_{{\bf k}}=\sqrt{\left(\xi^{-}_{{\bf k}}\right)^{2}+\overline{m}^{2}}$,  
$\xi_{{\bf k}}^{\pm}=(\xi_{{\bf k}}\pm\xi_{{\bf k}+{\bf Q}})/2$, 
$d_{\vk}=\cos k_x - \cos k_y$, 
and $N$ is the total number of lattice sites. 

The longitudinal spin susceptibility $\chi_{n (\mu)}$ is defined as  
$\chi_{n (\mu)} = \lim_{h \rightarrow 0} \left. \frac{\partial m}{\partial h} \right|_{n(\mu)}$  
for a fixed electron density (chemical potential). 
As clarified in Ref.~\onlinecite{kuboki17}, it is crucially important to specify which one is fixed 
inside a magnetic phase, the density or the chemical potential, because $\chi_{\mu}$ ($\chi_n$) does not describe 
the spin susceptibility of a system with a fixed density 
(chemical potential) even if the chemical potential (density) is tuned to reproduce the correct density (chemical potential); 
see also \eq{thermodynamics}. 
Below we focus on $\chi_n$  and results for $\chi_\mu$ are left to Appendix~C. 

As discussed in Ref.~\onlinecite{kuboki17}, calculations of $\chi_n$ are nontrivial already in the 
RPA because the chemical potential acquires a linear term 
in $h$ in a magnetic phase for a fixed density. Moreover, the superconducting gap is also expected 
to acquire a linear term in $h$ in a coexistence phase. Recalling that the RPA is equivalent to a mean-field approximation, 
we compute $\chi_n$ from the self-consistency equations [Eqs.~(\ref{self-n})-(\ref{self-m})] 
by taking a derivative with respect to $h$. 
We obtain the following matrix equation after taking the limit of $h\rightarrow 0$: 
\be
\left(
 \begin{array}{ccc}
a_{11} & a_{12} & a_{13} \\
a_{21} & a_{22} & a_{23} \\
a_{31} & a_{32} & a_{33} 
\end{array}
\right) 
\left(
\begin{array}{c}
\frac{\partial \mu}{\partial h} \\
\frac{\partial \Delta_{0}}{\partial h} \\
\frac{\partial \overline{m}} {\partial h} 
\end{array} 
\right)
=
\left(
\begin{array}{c}
0 \\
0 \\ 
\chi_{n}
\end{array} 
\right) \,.
\label{chin-eq}
\ee
The expressions for $a_{ij}$ are given in Appendix~A. 
Introducing the cofactors of the above matrix $A$ as $\tilde{a}_{ij}$ and 
noting $\partial \overline{m} /\partial h = 2V_m \chi_n +1/2$, 
we get the analytical expressions 
\be
\left(
\begin{array}{c}
\frac{\partial \mu}{\partial h} \\
\frac{\partial \Delta_{0}}{\partial h} \\
\chi_{n}
\end{array} 
\right)
=
\frac{\chi_{n}^{0}}{1-4V_m \chi_{n}^{0}} 
\left(
\begin{array}{c}
\tilde{a}_{31} /{\rm det} A \\
\tilde{a}_{32} /{\rm det} A \\
1
\end{array} 
\right) \,,
\label{chin-sol}
\ee
with $\chi_{n}^{0} = {\rm det} A /( 2 \tilde{a}_{33})$.  
Note that both $\mu$ and $\Delta_0$ can acquire a linear term in $h$, which is crucially important 
inside a magnetic phase as we shall show below.

\section{Results}
\begin{figure}[ht]
\centering
\includegraphics[width=8cm]{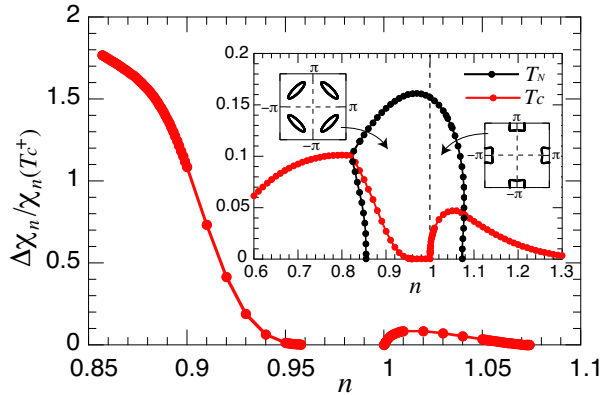}
\caption{(Color online) 
Density dependence of the jump of $\chi_{n}$ ($\Delta\chi_n$) along the $T_c$ curve  
inside the magnetic phase. The magnitude of the jump is scaled by $\chi_n$ just above $T_c$. 
Superconducting instability does not occur down to $T=0.0003$ in $0.96 < n < 1$, 
where $\Delta \chi_n$ cannot be defined. 
The inset shows the doping dependence of $T_c$ and $T_N$ 
by assuming a continuous phase transition and also depicts 
typical Fermi-surface pockets inside the 
magnetic phase on the hole- and electron-doped side. 
}
\label{chin-jump}
\end{figure}

We choose the band parameters $t'/t=-0.14$ and $t''/t=0.07$, which reproduce a hole-like Fermi surface 
typical to cuprate superconductors \cite{damascelli03}. For the interaction strength, 
we take $V_s /V_m =-3/8$ ($V_s <0$) so that the coexistence occurs at a reasonable temperature 
inside a magnetic phase. 
If the ratio of $| V_s | /V_m$ is too large (small), the superconductivity (magnetism) would become dominant.
We tune a value of $t/V_m$ to make sure that the magnetic phase is realized around the electron density 1 
with a domelike shape and we choose $t/V_m=0.8$; see the inset in \fig{chin-jump} for the phase diagram 
although details of the phase diagram are not important to our conclusions. 
We use $V_m$ as our unit of energy.

\begin{figure}[ht]
\centering
\includegraphics[width=14cm]{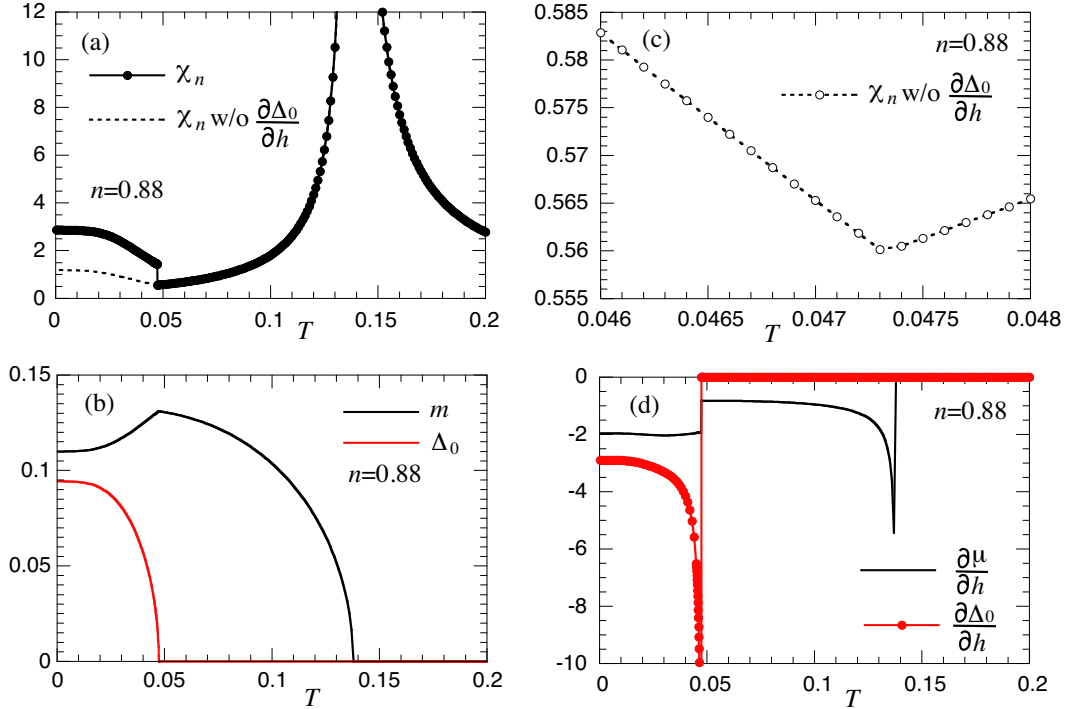}
\caption{(Color online) 
(a) Temperature dependence of the longitudinal spin susceptibility $\chi_n$ for a fixed density $n=0.88$. 
Antiferromagnetic and superconducting instabilities occur at $T_N =0.138$ and $T_c =0.0473$, respectively, 
and the microscopic coexistence of both orders is realized below $T_c$. 
The dotted line is $\chi_n$ when the contribution of $\frac{\partial \Delta_{0}} {\partial h}$ is neglected; 
"w/o" stands for "without". 
(b) Temperature dependence of the antiferromagnetic ($m$) and superconducting ($\Delta_0$) orders.  
(c) Enlarged view of the dotted line in (a) around $T_c$. 
(d) Temperature dependences of $\frac{\partial \mu} {\partial h}$ and $\frac{\partial \Delta_{0}} {\partial h}$, 
which diverge at $T_N$ and $T_c$, respectively, only on the low temperature side. $\frac{\partial \mu} {\partial h}$ 
exhibit a jump at $T_c$.  
}
\label{chin-T}
\end{figure}

Figure~\ref{chin-T}(a) shows the temperature dependence of the longitudinal spin susceptibility $\chi_n$ 
for a fixed density. 
The system is in a metallic phase in a high temperature region. 
With decreasing temperature $T$, $\chi_n$ is enhanced and diverges at $T_N=0.138$, 
signaling the instability toward an antiferromagnetic phase. 
Inside the magnetic phase, $\chi_n$ is suppressed quickly by the development of the magnetic order; 
see also \fig{chin-T}(b). 
With decreasing $T$ further, the magnetic phase exhibits the superconducting instability 
at $T_c=0.0473$ via a continuous phase transition as shown in \fig{chin-T}(b). 
The magnetism competes with superconductivity and the magnetic moment is suppressed by 
the superconductivity, but their microscopic coexistence is realized in $T<T_c$ as seen in various studies 
\cite{giamarchi91,inaba96,himeda99,sushkov04,yamase04a,inui88,lichtenstein00,senechal05,aichhorn06,capone06,reiss07,kancharla08,jwang14,bxzheng16,yamase16,kato88}. 
It might seem counterintuitive that the spin susceptibility is enhanced in the coexistence in \fig{chin-T}(a) 
because the spin degree of freedom tends to disappear by forming singlet pairings of electrons 
in the coexistence phase. However, given that the suppression of the spin susceptibility in $T<T_N$ is 
due to the development of the magnetic order,  it is reasonable that 
the suppression of the magnetic order by superconductivity yields the enhancement of the spin susceptibility 
in the coexistence.  

The major finding of the present work is that the spin susceptibility exhibits a jump at $T_c$. 
If we neglect a contribution from $\frac{\partial \Delta_0}{\partial h}$, the susceptibility 
shows a cusp at $T_c$ as illustrated in Figs.~\ref{chin-T}(a) and (c). 
The origin of the jump therefore lies in the emergence of $\frac{\partial \Delta_0}{\partial h}$ 
inside the magnetic phase. 

To understand the emergence of linear terms in $h$ inside the magnetic phase, 
we show in \fig{chin-T}(d) the temperature dependences of $\frac{\partial \mu}{\partial h}$ and 
$\frac{\partial \Delta_0}{\partial h}$. 
$\frac{\partial \mu}{\partial h}$ is zero in $T> T_N$, diverges at $T_N$ with $(T_N - T)^{-1/2}$ 
(Ref.~\onlinecite{misc-exponentTn}) only on the low temperature side, 
and becomes finite in $T< T_N$. 
As clarified in Ref.~\onlinecite{kuboki17}, 
the emergence of $\frac{\partial \mu}{\partial h}$ comes from two factors: 
the breaking of spin rotational symmetry in the magnetic phase and 
an infinitesimally small magnetic field $h$ applied along the same direction as the magnetic moment. 
It is the term of $\frac{\partial \mu}{\partial h}$ that ensures the reasonable suppression of $\chi_n$ 
with decreasing $T$ inside the magnetic phase \cite{kuboki17}.  
Similarly, a contribution from $\frac{\partial \Delta_0}{\partial h}$ should also be taken into account, 
but this quantity is zero until the superconducting order parameter starts to develop. 
At $T_c$, $\frac{\partial \Delta_0}{\partial h}$ diverges with $(T_c - T)^{-1/2}$ only on the side of $T< T_c$ and 
becomes finite at lower $T$. This singular behavior $\frac{\partial \Delta_0}{\partial h}$ at $T_c$ leads to 
the jump of the spin susceptibility in \fig{chin-T}(a). 

Analytical understanding is obtained by studying the asymptotic behavior of each matrix element 
in \eq{chin-eq} in the vicinity of $T_c$, which yields $a_{12} \propto \Delta_0$, $a_{21} \propto \Delta_0$, 
$a_{22} \propto \Delta_0^2$, $a_{23} \propto \Delta_0$, $a_{32} \propto \Delta_0$, and a finite value 
for the other elements; see Appendix~B for $a_{22} \propto \Delta_0^2$. 
Simple algebra then shows that 
${\rm det} A \approx  {\rm det}A'' \Delta_0^{2}$, 
$\tilde{a}_{31} \approx \tilde{a}_{31}'' \Delta_0^{2}$, 
$\tilde{a}_{32} \approx \tilde{a}_{32}' \Delta_0$, and 
$\tilde{a}_{33} \approx \tilde{a}_{33}'' \Delta_0^{2}$, where 
$ {\rm det}A''$,  $\tilde{a}_{31}''$, $\tilde{a}_{32}'$, and $\tilde{a}_{33}''$ are 
finite at $T_c$. Therefore \eq{chin-sol} can be written close to $T_c$ as 
\be
 \left(
\begin{array}{c}
\frac{\partial \mu}{\partial h} \\
\frac{\partial \Delta_{0}}{\partial h} \\
\chi_{n}
\end{array} 
\right)
=
\frac{\chi_{n}^{0-}}{1-4V_m \chi_{n}^{0-}} 
\left(
\begin{array}{c}
\tilde{a}_{31}''/{\rm det} A'' \\
\tilde{a}_{32}' /{\rm det} A'' \times \Delta_0^{-1} \\
1
\end{array} 
\right) \,,
\label{chin-sol-Tc-}
\ee
and 
\be
 \chi_{n}^{0-} = \frac{1}{2} 
\frac{{\rm det} A^{''}} {\tilde{a}_{33}^{''}} \,.
\label{chin0-}
\ee
The factor of $\Delta_0^2$ is cancelled out except for 
$\frac{\partial \Delta_0}{\partial h}$. This is the reason why $\frac{\partial \Delta_0}{\partial h}$ 
shows a divergence of $(T_c-T)^{-\beta}$ with $\beta=1/2$ (Ref.~\onlinecite{misc-exponentTn}.)

The solution \eq{chin-sol-Tc-} is valid at $T=T_c^-$. 
In $T> T_c$ we have $a_{i2}=a_{2j}=0$ for $i,j=1,2,3$. Hence the matrix equation (\ref{chin-eq}) 
is reduced to a $2 \times 2$ matrix equation. It is then straightforward to obtain 
$\chi_{n}=\frac{\chi_{n}^{0+}}{1-4V_m \chi_{n}^{0+}}$ at $T=T_c^+$ with  
\begin{eqnarray}
\chi_{n}^{0+} = \frac{1}{2} 
\left( 
a_{33} - \frac{a_{13}a_{31}}{a_{11}} 
\right)\,. 
 \label{chin0+}
\end{eqnarray}
A comparison between Eqs.~(\ref{chin0-}) and (\ref{chin0+}) immediately indicates that 
$\chi_n^{0-}$ and $\chi_n^{0+}$ becomes different in general, because $\chi_{n}^{0-}$ encodes 
the effect of superconductivity through the coefficients 
of the quadratic term of superconducting order parameter 
whereas $\chi_{n}^{0+}$ is characterized by 
the quantities in the purely magnetic phase. 
This explains the reason why $\chi_n$ exhibits a jump at $T_c$. 
Similarly, we can understand the reason why $\frac{\partial \mu}{\partial h}$ exhibits a jump at $T_c$ 
in \fig{chin-T}(b).

The magnitude of the susceptibility jump may be denoted as $\Delta \chi_n$. The jump is then quantified by considering 
the ratio of $\Delta \chi_n$ to $\chi_n$ at $T=T_c^+$, which becomes dimensionless. 
In \fig{chin-T}(a) the ratio is around 1.5. 
This ratio is not universal and depends on the density. 
We compute the ratio of 
$\Delta \chi_n / \chi_n(T_c^+)$ along the $T_c$ curve inside the magnetic phase in \fig{chin-jump}. 
The ratio becomes as large as 1.8 on the hole-doped side $(n<1)$ and 
rather small around 10~\% at most on the electron-doped side $(n>1)$. 
This strong asymmetry with respect to $n=1$ originates from a presence of $t'$ and $t''$, which 
breaks particle-hole symmetry. At $n=1$, the model exhibits an insulating state and a metallic state 
is realized upon carrier doping. The Fermi-surface pockets in the antiferromagnetic phase are then 
realized around $\vk=(\pi/2,\pi/2)$ and $(\pi,0)$ in the hole- and electron-doped case, respectively, 
as shown in the inset in \fig{chin-jump}. 
Hence the $d$-wave superconductivity can develop rather quickly 
on the electron-doped side, whereas it is strongly suppressed by the magnetism close to $n=1$ and 
increases rather gradually upon further hole doping because of the expansion of the Fermi-surface pocket  
toward the directions of $(\pi,0)$ and $(0,\pi)$ where the $d$-wave superconductivity acquires larger energy gain. 
This availability of states around $(\pi,0)$ and $(0,\pi)$ 
also explains the reason why the $T_c$ curve is smooth on the electron-doped side 
when entering the magnetic phase whereas it is suppressed strongly 
on the hole-doped side. 
Therefore, because of the difference of the underlying Fermi-surface pockets, 
the magnetism tends to be suppressed more substantially by the superconductivity  
on the hole-doped side as seen in \fig{chin-T}(b), which then yields the enhancement of  
the ratio $\Delta \chi_n / \chi_n(T_c^+)$ as shown in \fig{chin-jump}. 
While we have focused on $\chi_n$ here, we find that 
a jump of $\chi_{\mu}$ is much more pronounced as presented in \fig{chimu-T} in Appendix~C. 

The thermodynamic relation \eq{thermodynamics} was derived in Ref.~\onlinecite{kuboki17} where 
a pure magnetic phase was considered. 
While the additional term $\frac{\partial \Delta_{0}}{\partial h}$ emerges in the superconducting state, 
we obtain the same expression as \eq{thermodynamics} even in the coexistence 
on the basis of thermodynamics \cite{callen} alone.  
The effect of superconductivity enters as a {\it total} derivative in the sense that the $h$ dependence 
of $\Delta_0$ is considered when evaluating $\chi_{n (\mu)}$, $\left. \frac{\partial n}{\partial h} \right|_{\mu}$, and 
$\left. \frac{\partial \mu}{\partial h} \right|_{n}$. 
While the thermodynamic principle yields \eq{thermodynamics}, 
it is highly nontrivial to infer \eq{thermodynamics} even in a simple mean-field theory. 
We can however check it explicitly in a whole temperature region 
including the region near $T_c$ in \fig{chin-T} 
by evaluating also $\chi_{\mu}$ and $\frac{\partial n}{\partial h}$ [see \eq{chimu-sol}] numerically.

\section{Discussions and conclusions} 
The jump of the longitudinal spin susceptibility in \fig{chin-T}(a) might look similar to the jump of 
the specific heat at $T_c$ in the BCS theory \cite{bardeen57}, but the underlying mechanism is different. 
First, the jump of $\chi_{n (\mu)}$ is obtained when the superconducting instability occurs 
inside a magnetic phase, not in the normal phase. 
Second, the jump of the specific heat comes from a typical feature of mean-field theory, where $\Delta_{0}^{2}$ is 
proportional to $T_c-T$ in $T < T_c$ (Ref.~\onlinecite{bardeen57}).   
However, a term of $\Delta_{0}^{2}$ is cancelled out in the spin 
susceptibility as we have clarified in \eq{chin-sol-Tc-}. 


The jump of $\chi_{n (\mu)}$ originates from the emergence of $\frac{\partial \Delta_0}{\partial h}$ 
below $T_c$ and its singular behavior at $T_c$ as we have already explained. 
The emergence of $\frac{\partial \Delta_0}{\partial h}$ itself traces back to 
the breaking of spin rotational symmetry inside a magnetic phase \cite{kuboki17} and thus 
is a general feature independent of details of models and approximations. 
Recalling that the superconductivity occurs via a continuous phase transition, 
it is likely a general feature that $\frac{\partial \Delta_0}{\partial h}$ exhibits 
a power-law divergence at $T_c$ even beyond the present mean-field theory 
as long as a continuous phase transition into the coexistence survives. 
The longitudinal spin susceptibility couples to such singular behavior of $\frac{\partial \Delta_0}{\partial h}$ and 
thus should also be characterized by a certain singularity at $T_c$. Physically $\chi_{n(\mu)}$ cannot show 
a divergence at $T_c$, otherwise the magnetic instability would occur at $T_c$. 
Hence the only possible singularity of the longitudinal spin susceptibility is a jump at $T_c$. 
One might wonder about a possible cusp at $T_c$. In this case, the spin susceptibility 
would become continuous across $T_c$, which is unlikely in general because of the fact that 
$\frac{\partial \Delta_0}{\partial h}$ shows the singularity only on the low temperature side of $T_c$. 
Therefore we believe that the jump of the longitudinal spin susceptibility can be a general feature 
when the superconducting instability occurs via a continuous phase transition inside a magnetic phase 
and the coexistence is realized at lower temperatures. 

It is important to recognize that the above argument relies on only two general features inside the 
magnetic phase: the emergence of $\frac{\partial \Delta_0}{\partial h}$ and its singular behavior associated with 
a continuous phase transition of superconductivity.  
The underlying magnetic structure, namely a value of $\vQ$ does not matter.  
Furthermore, the argument does not rely on the symmetry of superconductivity as long as the superconductivity 
coexists with the magnetism.

Recently direct evidence of the coexistence of superconductivity and magnetism was reported for 
multilayer cuprates \cite{mukuda12,kunisada20}. Hence it is interesting to test a jump of the longitudinal spin susceptibility 
with a wavevector $\vQ=(\pi,\pi)$ by spin-spin relaxation time in NMR or polarized neutron scattering measurements more directly. 
While electron correlations specific to cuprates are not included in the present theory, 
the present minimal model may be regarded as an effective one containing correlations effects 
via model parameters, for example, a model obtained 
after the slave-boson mean-field approximation to the $t$-$J$ model \cite{inaba96,yamase04a} or 
a low-energy effective model obtained after integrating high-energy degrees  of freedom \cite{reiss07,jwang14,yamase16}.

Higher-order corrections not included in the present theory may modify quantitative aspects. 
The ordering tendency of both magnetism and superconductivity would be suppressed. 
However, the present theory is still applicable as long as the coexistence survives. 
The exponent of the singularity of $\frac{\partial \Delta_0}{\partial h}$ at $T_c$ would be changed from 
the mean-field value. This effect is not expected to generate a singularity different from 
a jump in the longitudinal spin susceptibility as we have already argued above. 
The magnitude of the jump may be suppressed or enhanced by the effect beyond the RPA. 
Since longitudinal spin fluctuations are already suppressed inside a magnetic phase 
owing to the development of the magnetic order, the jump of the spin susceptibility may not change drastically 
from the present mean-field theory as long as the coexistence remains and the magnetism is 
suppressed by the onset of superconductivity.

The analytical expressions are different between $\chi_{n}$ [\eq{chin-sol}]  
and $\chi_{\mu}$ [see \eq{chimu-sol}]. Which quantity should be 
employed when discussing actual materials?  
The electron density is usually fixed for actual materials and thus $\chi_n$ seems more appropriate.  
However, the situation may not be so trivial in several cases. 
For example, for multilayer cuprates \cite{mukuda12,kunisada20}, each CuO$_2$ plane can be regarded 
as being in contact with a charge reservoir because there is a charge transfer among different CuO$_2$ planes 
inside the unit cell and the coexistence in question is realized only in a certain CuO$_2$ plane among them.  
Another example is a system where several bands cross the Fermi energy. 
If a certain band contributes to the spin susceptibility substantially more than the others, 
the other bands are regarded as spectators. Those systems may be modeled by employing 
an effective one-band model for a fixed chemical potential. 
In this case, a jump of the spin susceptibility tends to be much more enhanced; 
see \fig{chimu-T} in Appendix~C.

In experiments, it is not easy to distinguish between microscopic  coexistence and phase separation, 
which can be frequently controversial for many materials. 
The present work casts a light on this issue, because a jump of the longitudinal spin susceptibility 
can be utilized as a thermodynamic probe of the microscopic coexistence of superconductivity and magnetism. 
The magnitude of the jump depends on details of the system. 
In the present model, we find that the jump is typically pronounced when 
the magnetization is fairly suppressed by the superconducting order [\fig{chin-T}(b)] 
as seen in the hole-doped region in \fig{chin-jump}. 
This suggests that a material, in which superconductivity and magnetism compete with each other, 
but either one does not become dominant, 
is suitable to test the present theory, for example, iron-based superconductors \cite{stewart11} 
as well as multilayer cuprates \cite{mukuda12,kunisada20}. 
See also results of $\chi_{\mu}$ [\fig{chimu-T}(b)] in Appendix~C. 
On the other hand, if superconductivity and magnetism are phase-separated inside a material 
via a continuous phase transition, 
the longitudinal spin susceptibility does not exhibit a jump at $T_c$. 

The jump of the longitudinal spin susceptibility at superconducting instability inside a magnetic phase 
is a fundamental feature associated with superconductivity. 
It is curious that we cannot find the corresponding anomaly in the 110-year history of superconductivity. 
There seem several reasons. First, 
it is rather recently that a possible coexistence of superconductivity and magnetism was reported in 
various materials \cite{stewart11, pfleiderer09,mukuda12,kunisada20}. 
Second, the longitudinal susceptibility in question is not necessarily at $\vQ=(0,0)$. 
Hence an experimental effort may not be made without a theoretical input. 

One might wonder about a situation in which superconductivity first sets in and magnetic instability occurs later 
inside the superconducting phase. 
In this case, nothing special is expected in the longitudinal spin susceptibility: 
it would show a cusp at $T_c$ and a power-law divergence at $T_N (< T_c)$. 
The present theory is applicable to the case that superconducting instability occurs 
inside a magnetic phase.

In summary, we reveal that the longitudinal spin susceptibility exhibits 
a new anomaly associated with superconducting instability inside a magnetic phase, 
namely the jump of the susceptibility. This anomaly originates from the emergence of 
$\frac{\partial \Delta_0}{\partial h}$ in $T< T_c$, whose origin traces back to the breaking of spin rotational 
symmetry inside the magnetic phase. The jump of the longitudinal spin susceptibility 
is a thermodynamic signature of the microscopic coexistence 
of superconductivity and magnetism, and can be tested in various materials. 
While our calculations are performed in mean-field theory,  
it can be argued that the jump of the longitudinal spin susceptibility reflects a general feature associated with 
superconducting instability inside a magnetically ordered phase.

\acknowledgments
The authors thank P. Jakubczyk for a critical reading of the manuscript and thoughtful comments, and 
K. Kuboki for valuable discussions throughout the present project. 
They also thank D. Aoki, M. Fujita, H.  Mukuda, and T. Terashima for helpful comments from an experimental point of view.  
This work was supported by JSPS KAKENHI Grants No.~JP20H01856 and JST-Mirai Program Grant Number JPMJMI18A3, Japan. 



\appendix
\section{Matrix elements for a fixed density} 
The matrix elements in \eq{chin-eq} are computed by taking a derivative of the self-consistency 
equations [Eqs.~(\ref{self-n})-(\ref{self-m})] with respect to a field $h$: 
\bea
&&a_{11}=\frac{1}{N}\sideset{}{'}\sum_{\vk} 
\left(   \frac{(\eta_{\vk}^{+})^{2} g^{+}_{\vk} + \tanh  \frac{\lambda_{{\bf k}}^{+}}{2T} } {\lambda_{\vk}^{+}} 
+ \frac{(\eta_{\vk}^{-})^{2} g^{-}_{\vk} + \tanh  \frac{\lambda_{{\bf k}}^{-}}{2T} } {\lambda_{\vk}^{-}} 
\right)  \,, \\
&&a_{12}= -\frac{2V_s}{N}\sideset{}{'}\sum_{\vk} \Delta_{\vk} d_{\vk} 
\left(  \frac{\eta^{+}_{\vk} g^{+}_{\vk}} {\lambda_{\vk}^{+}} + \frac{\eta^{-}_{\vk} g^{-}_{\vk}} {\lambda_{\vk}^{-}} 
\right) \label{a12} \,, \\
&& a_{13} = - \frac{2m V_m}{N}\sideset{}{'}\sum_{\vk} \frac{1}{D_{\vk}} 
\left(   \frac{(\eta_{\vk}^{+})^{2} g^{+}_{\vk} + \tanh  \frac{\lambda_{{\bf k}}^{+}}{2T} } {\lambda_{\vk}^{+}} 
- \frac{(\eta_{\vk}^{-})^{2} g^{-}_{\vk} + \tanh  \frac{\lambda_{{\bf k}}^{-}}{2T} } {\lambda_{\vk}^{-}} 
\right) \,, \\
&& a_{21}= -\frac{1}{4V_s} a_{12} \label{a21} \,, \\
&& a_{22} = -1 - \frac{V_s}{N}\sideset{}{'}\sum_{\vk} d_{\vk}^{2} 
\left(  \frac{\Delta_{\vk}^{2} g^{+}_{\vk} + \tanh  \frac{\lambda_{{\bf k}}^{+}}{2T} } {\lambda_{\vk}^{+}} 
+ \frac{\Delta_{\vk}^{2} g^{-}_{\vk} + \tanh  \frac{\lambda_{{\bf k}}^{-}}{2T} } {\lambda_{\vk}^{-}}  
\right) \,, \\
&&a_{23} =  -\frac{mV_m}{N}\sideset{}{'}\sum_{\vk} \frac{\Delta_{\vk} d_{\vk}}{D_{\vk}}
\left(  \frac{\eta^{+}_{\vk} g^{+}_{\vk}} {\lambda_{\vk}^{+}} - \frac{\eta^{-}_{\vk} g^{-}_{\vk}} {\lambda_{\vk}^{-}} 
\right) \label{a23} \,, \\
&& a_{31} = \frac{1}{2} a_{13} \,, \\
&& a_{32} = - 2 V_s a_{23} \label{a32}\,, \\
&& a_{33} =   \frac{1}{2N}\sideset{}{'}\sum_{\vk} \frac{(\xi_{\vk}^{-})^{2}}{D_{\vk}^{3}} 
\left(   \frac{\eta_{\vk}^{+}}{\lambda_{\vk}^{+}} \tanh  \frac{\lambda_{{\bf k}}^{+}}{2T} 
- \frac{\eta_{\vk}^{-}}{\lambda_{\vk}^{-}} \tanh  \frac{\lambda_{{\bf k}}^{-}}{2T} 
\right) \\
&& \hspace{18mm}
+ \frac{(2m V_m)^2}{2N}\sideset{}{'}\sum_{\vk} \frac{1}{D_{\vk}^2} 
\left(   \frac{(\eta_{\vk}^{+})^{2} g^{+}_{\vk} + \tanh  \frac{\lambda_{{\bf k}}^{+}}{2T} } {\lambda_{\vk}^{+}} 
+ \frac{(\eta_{\vk}^{-})^{2} g^{-}_{\vk} + \tanh  \frac{\lambda_{{\bf k}}^{-}}{2T} } {\lambda_{\vk}^{-}} 
\right) \,.
\eea
Here $g^{\pm}_{\vk}$ is given by 
\be
g_{\vk}^{\pm} = - \frac{1}{(\lambda_{\vk}^{\pm})^2} \tanh \frac{\lambda_{\vk}^{\pm}} {2T} 
+ \frac{1}{2T} \frac{1}{\lambda_{\vk}^{\pm}} \frac{1}{\cosh^2 \frac{\lambda_{\vk}^{\pm}}{2T}} \,.
\ee

\section{Asymptotic analysis near $\boldsymbol{T_c}$} 
A matrix element $a_{22}$ is proportional to $\Delta_{0}^2$ in the vicinity of $T_c$ as we mention in the main text. 
Here we provide the outline of the derivation. 

The element $a_{22}$ depends on $T, \mu, \Delta_{0}$, and $m$ when $n$ is fixed. 
Here $\mu, \Delta_{0}$, and $m$ also depend on $T$. Hence we expand $a_{22}$ with respect to $T$ around $T_c$
\bea
&&a_{22} (T) = a_{22}(T_c) + \left. \frac{{\rm d} a_{22}}{{\rm d} T}\right| _{T_c}  (T-T_c) + \cdots \,,\\
&& \hspace{12mm} =  \left. \frac{{\rm d} a_{22}}{{\rm d} T}\right| _{T_c}  (T-T_c) + \cdots 
\label{a22-T}\,.
\eea
Note that \eq{self-D} certifies $a_{22}=0$ at $T=T_c^{-}$. 
It is cumbersome to compute $\frac{{\rm d} a_{22}}{{\rm d} T}$, which is given by 
\be
\frac{{\rm d} a_{22}}{{\rm d} T} = \left(
\frac{\partial }{\partial T} + \frac{{\rm d} \mu}{{\rm d} T} \frac{\partial }{\partial \mu} 
+\frac{{\rm d} \Delta_{0}^2}{{\rm d} T} \frac{\partial }{\partial \Delta_{0}^{2}} 
+\frac{{\rm d} m}{{\rm d} T} \frac{\partial }{\partial m} 
\right)a_{22} \,.
\label{diffa22T}
\ee
Here we have considered that $a_{22}$ is a function of $\Delta_{0}^2$ so that 
$\frac{{\rm d} \Delta_{0}^2}{{\rm d} T}$ becomes regular at $T_c$. 
It is straightforward to obtain 
\bea
&& \left.\frac{\partial a_{22}}{\partial T}\right|_{T_c} = \frac{V_s} {2T_c^2 N} \sideset{}{'}\sum_{\vk} 
d_{\vk}^2 \left(
\frac{1}{\cosh^2 \frac{\eta_{\vk}^{+}}{2T_c}} + \frac{1}{\cosh^2 \frac{\eta_{\vk}^{-}}{2T_c}}
\right) \,, \label{a22T}\\
&& \left.\frac{\partial a_{22}}{\partial \mu}\right|_{T_c} = -\frac{V_s} {N} \sideset{}{'}\sum_{\vk} 
d_{\vk}^2 \left[
\frac{1}{(\eta_{\vk}^{+})^{2}} \tanh \frac{\eta_{\vk}^{+}}{2T_c} - 
  \frac{1}{2T_c} \frac{1}{\eta_{\vk}^{+}} \frac{1}{\cosh^2 \frac{\eta_{\vk}^{+}}{2T_c}} \right. \nonumber\\
&&\left. \hspace{42mm}  + \frac{1}{(\eta_{\vk}^{-})^{2}} \tanh \frac{\eta_{\vk}^{-}}{2T_c} - 
  \frac{1}{2T_c} \frac{1}{\eta_{\vk}^{-}} \frac{1}{\cosh^2 \frac{\eta_{\vk}^{-}}{2T_c}} 
\right] \,, \\
&& \left.\frac{\partial a_{22}}{\partial \Delta_{0}^{2}}\right|_{T_c} = \frac{6V_s^3} {N} \sideset{}{'}\sum_{\vk} 
d_{\vk}^4 \left[
\frac{1}{(\eta_{\vk}^{+})^{3}} \tanh \frac{\eta_{\vk}^{+}}{2T_c} - 
  \frac{1}{2T_c} \frac{1}{(\eta_{\vk}^{+})^2} \frac{1}{\cosh^2 \frac{\eta_{\vk}^{+}}{2T_c}} \right. \nonumber\\
&&\left. \hspace{45mm}  + \frac{1}{(\eta_{\vk}^{-})^{3}} \tanh \frac{\eta_{\vk}^{-}}{2T_c} - 
  \frac{1}{2T_c} \frac{1}{(\eta_{\vk}^{-})^2} \frac{1}{\cosh^2 \frac{\eta_{\vk}^{-}}{2T_c}} 
\right] \,, \\
&& \left.\frac{\partial a_{22}}{\partial m}\right|_{T_c} = \frac{4m V_m^2 V_s} {N} \sideset{}{'}\sum_{\vk} 
d_{\vk}^2 \frac{1}{D_{\vk}} 
\left[
\frac{1}{(\eta_{\vk}^{+})^{2}} \tanh \frac{\eta_{\vk}^{+}}{2T_c} - 
  \frac{1}{2T_c} \frac{1}{\eta_{\vk}^{+}} \frac{1}{\cosh^2 \frac{\eta_{\vk}^{+}}{2T_c}} \right. \nonumber\\
&&\left. \hspace{52mm}  - \frac{1}{(\eta_{\vk}^{-})^{2}} \tanh \frac{\eta_{\vk}^{-}}{2T_c} +  
  \frac{1}{2T_c} \frac{1}{\eta_{\vk}^{-}} \frac{1}{\cosh^2 \frac{\eta_{\vk}^{-}}{2T_c}} 
\right] \label{a22m} \,. 
\eea
To calculate $\frac{{\rm d} \mu}{{\rm d} T}$, $\frac{{\rm d} \Delta_{0}^2}{{\rm d} T}$, and 
$\frac{{\rm d} m}{{\rm d} T}$, we take a derivative of each self-consistency equation 
[Eqs.~(\ref{self-n})-(\ref{self-m})] with respect to $T$, 
yielding the following matrix equation: 
\be
\left(
 \begin{array}{ccc}
a_{11} & a_{12} / (2\Delta_0) & 2 V_m a_{13} \\
a_{21} /(2\Delta_0)  & \overline{a}_{22}  & 2 V_m a_{23}  /(2\Delta_0) \\
a_{31} & a_{32}  / (2\Delta_0)  & 2 V_m a_{33} -1
\end{array}
\right) 
\left(
\begin{array}{c}
\frac{{\rm d} \mu}{{\rm d} T} \\
\frac{{\rm d} \Delta_{0}^2}{{\rm d} T} \\
\frac{{\rm d} m} {{\rm d} T} 
\end{array} 
\right)
=
\left(
\begin{array}{c}
c_1 \\
c_2 /(2 \Delta_{0})  \\ 
c_3
\end{array} 
\right) \,,
\label{diffT-eq}
\ee
where $a_{ij}$ are the same as those in \eq{chin-eq} and 
\bea
&& \overline{a}_{22} = - \frac{V_s^3}{N} \sideset{}{'}\sum_{\vk} d_{\vk}^4 \left( 
\frac{g_{\vk}^{+}}{\lambda_{\vk}^{+}} + \frac{g_{\vk}^{-}}{\lambda_{\vk}^{-}}
\right) \,, \\ 
&&c_1= - \frac{1}{2 T_c^2 N} \sideset{}{'}\sum_{\vk} 
\left(
\frac{\eta_{\vk}^{+}}{\cosh^2 \frac{\eta_{\vk}^{+}}{2T_c}} + \frac{\eta_{\vk}^{-}}{\cosh^2 \frac{\eta_{\vk}^{-}}{2T_c}}
\right) \,, \\
&&c_2= - \frac{1}{4 T_c^2 N} \sideset{}{'}\sum_{\vk} \Delta_{\vk} d_{\vk} 
\left(
\frac{1}{\cosh^2 \frac{\eta_{\vk}^{+}}{2T_c}} + \frac{1}{\cosh^2 \frac{\eta_{\vk}^{-}}{2T_c}}
\right) \,, \\
&&c_3= \frac{mV_m}{2 T_c^2 N} \sideset{}{'}\sum_{\vk} \frac{1}{D_{\vk}}  
\left(
\frac{\eta_{\vk}^{+}}{\cosh^2 \frac{\eta_{\vk}^{+}}{2T_c}} - \frac{\eta_{\vk}^{-}}{\cosh^2 \frac{\eta_{\vk}^{-}}{2T_c}}
\right) \,. 
\eea
Note that $a_{12}/(2\Delta_0)$, $a_{21}/(2\Delta_0)$, $a_{23}/(2\Delta_0)$, 
$a_{32}/(2\Delta_0)$, and $c_{2}/(2\Delta_0)$ 
become finite in the limit of $\Delta_{0} \rightarrow 0$. By solving \eq{diffT-eq} and 
evaluating Eqs.~(\ref{a22T})-(\ref{a22m}),  we obtain the quantity of 
$\frac{{\rm d} a_{22}}{{\rm d} T}$ given in \eq{diffa22T}. 

To write $a_{22}$ in terms of $\Delta_{0}$, we recall that $\Delta_0$ can be parameterized as 
$\Delta_0 = a (T_c -T)^{\frac{1}{2}}$ near $T_c$ in mean-field theory. Hence \eq{a22-T} can be written as 
\bea
&&a_{22} (T) = -\frac{1}{a^2} \left. \frac{{\rm d} a_{22}}{{\rm d} T}\right| _{T_c} \Delta_0^2 + \cdots \\
&&\hspace{13mm} = \left( \left. \frac{{\rm d} \Delta_{0}^2}{{\rm d} T}\right| _{T_c} \right)^{-1} 
\left. \frac{{\rm d} a_{22}}{{\rm d} T}\right| _{T_c} \Delta_0^2 + \cdots \,.
\eea
This coefficient of $\Delta_0^2$ is computed from Eqs.~(\ref{diffa22T}) and (\ref{diffT-eq}) numerically. 

The coefficients of proportionality of $a_{12} \propto \Delta_0$,   $a_{21} \propto \Delta_0$,  $a_{23} \propto \Delta_0$,  
and $a_{32} \propto \Delta_0$ are easily read off from Eqs.~(\ref{a12}), (\ref{a21}), (\ref{a23}), and (\ref{a32}), respectively. 
After simple algebra, we can evaluate numerically $\chi_{n}^{0-}$ [\eq{chin0-}] and other quantities in \eq{chin-sol-Tc-}.

\section{Spin susceptibility for a fixed chemical potential} 
We have presented results for a fixed density in the main text. Here we present 
key expressions and results for a fixed chemical potential. 

Under the condition of a fixed chemical potential, we take a derivative of each self-consistency 
equation [Eqs.~(\ref{self-n})-(\ref{self-m})] with respect to a field $h$ and then take the limit of 
$h \rightarrow 0$. We then obtain 
\be
\left(
 \begin{array}{ccc}
b_{11} & b_{12} & b_{13} \\
b_{21} & b_{22} & b_{23} \\
b_{31} & b_{32} & b_{33} 
\end{array}
\right) 
\left(
\begin{array}{c}
\frac{\partial n}{\partial h} \\
\frac{\partial \Delta_{0}}{\partial h} \\
\frac{\partial \overline{m}} {\partial h} 
\end{array} 
\right)
=
\left(
\begin{array}{c}
0 \\
0 \\ 
\chi_{\mu}
\end{array} 
\right) \,.
\label{chimu-eq}
\ee
The corresponding expression for a fixed density is given in \eq{chin-eq}, where 
$\frac{\partial \mu}{\partial h}$ appears instead of $\frac{\partial n}{\partial h}$. 
The matrix elements $b_{ij}$ are the same as $a_{ij}$ 
except for $b_{11}=-1$ and $b_{21}=b_{31}=0$. 
We describe the $(ij)$-th cofactor of the above matrix $B$ as $\tilde{b}_{ij}$. 
The analytical solution is given by 
\be
\left(
\begin{array}{c}
\frac{\partial n}{\partial h} \\
\frac{\partial \Delta_{0}}{\partial h} \\
\chi_{\mu}
\end{array} 
\right)
=
\frac{\chi_{\mu}^{0}}{1-4V_m \chi_{\mu}^{0}} 
\left(
\begin{array}{c}
\tilde{b}_{31} /{\rm det} B \\
\tilde{b}_{32} /{\rm det} B \\
1
\end{array} 
\right) \,,
\label{chimu-sol}
\ee
where $\chi_{\mu}^{0} = {\rm det} B / (2\tilde{b}_{33})$. 
Note that the functional form of $\chi_{\mu}$ is exactly the same as $\chi_n$ in \eq{chin-sol}, 
but the matrix $B$ is not the same as $A$. Hence the resulting values of 
$\frac{\partial \Delta_{0}}{\partial h}$, 
$\chi_{\mu}$, and $\chi_{\mu}^{0}$ become different from those for a fixed density 
inside a magnetic phase. No difference occurs in the normal phase because of 
vanishing additional linear contributions in $h$, 
i.e., $\frac{\partial n}{\partial h}=\frac{\partial \Delta_{0}}{\partial h}=0$.  

\begin{figure}[ht]
\centering
\includegraphics[width=15cm]{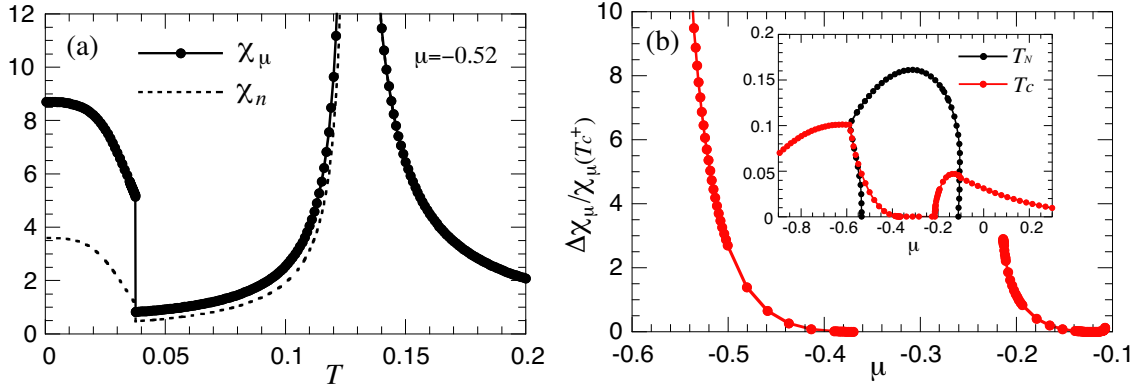}
\caption{(Color online)
(a) Temperature dependence of the longitudinal spin susceptibility $\chi_{\mu}$ for a fixed chemical potential $\mu=-0.52$. 
Antiferromagnetic and superconducting instabilities occur at $T_N =0.127$ and $T_c =0.0375$, respectively, 
and the microscopic coexistence is realized below $T_c$. 
$\chi_n$ (dotted line) is obtained by tuning the density to reproduce $\mu=-0.52$ at each temperature 
and becomes different from $\chi_{\mu}$ below $T_N$. $\chi_{\mu}$ and $\chi_n$ are connected with each other 
via the thermodynamic relation \eq{thermodynamics}. 
(b) Chemical potential dependence of the jump of $\chi_{\mu}$ ($\Delta\chi_{\mu}$) along the curve of 
$T_c$ inside the magnetic phase; the magnitude of the jump is scaled by $\chi_{\mu}$ just above $T_c$. 
The inset shows the phase diagram in the plane of the chemical potential and  temperature. 
Superconducting instability does not occur down to $T=0.0003$ in $-0.37  < \mu < -0.214$.   
The hole-doped region is in $\mu < -0.214$ and the electron-doped region is in $\mu > -0.214$. 
}
\label{chimu-T}
\end{figure}

For completeness, we present in \fig{chimu-T}(a) temperature dependence of $\chi_{\mu}$ 
for $\mu=-0.52$; the density is $n=0.876$ at $T=0$ and thus may be reasonably compared with \fig{chin-T}. 
Similar to \fig{chin-T}(a), $\chi_{\mu}$ exhibits a jump at $T_c=0.0375$.  
The anomaly at $T_c$ is pronounced much more than $\chi_n$ shown in \fig{chin-T}(a). 
This pronounced anomaly is also highlighted by comparing  
the doping, i.e., chemical potential, dependence of the jump [\fig{chimu-T}(b)] with that of $\chi_n$ (\fig{chin-jump}). 
While the jump might seem small on the electron-doped side in \fig{chimu-T}(b), 
its magnitude is comparable to that in the hole-doped region in \fig{chin-jump}. 
Besides the amplitude of the jump, another marked difference between \fig{chimu-T}(b) and \fig{chin-jump} lies 
in a small doping region on the electron-doped side:  
the clear enhancement of $\Delta \chi_{\mu} / \chi_{\mu}(T_c^+)$ close to $\mu=-0.214$ in \fig{chimu-T}(b), but 
its suppression close to $n=1$ in \fig{chin-jump}. 
The enhancement of $\Delta \chi_{\mu} / \chi_{\mu}(T_c^+)$ here should be understood with a special care. 
First, it does not imply behavior of a divergence. 
Since the magnetic susceptibility should not diverge at $T_c$ inside the magnetic phase, 
$\Delta \chi_{\mu}$ should be finite. In addition, $\chi_{\mu}(T_c^+)$ becomes small deeply inside the magnetic phase, 
but retains a finite value at $T=0$; this is true even at half-filling in the present model. 
Second, we checked numerically that $\chi_{\mu}$ at $T=T_c^+$ decreases upon approaching $\mu=-0.214$ 
from the above, but the jump of $\chi_{\mu}$, namely $\Delta \chi_{\mu}$ stays roughly around $0.75$. 
Hence the enhancement of $\Delta \chi_{\mu} / \chi_{\mu}(T_c^+)$ in \fig{chimu-T}(b) does not mean the 
enhancement of $\Delta \chi_{\mu}$, but comes from the suppression of $\chi_{\mu}(T_c^+)$.

In \fig{chimu-T}(a) we also plot $\chi_{n}$ by tuning the density to reproduce 
the correct chemical potential $\mu=-0.52$ at each temperature. 
$\chi_{n}$ and $\chi_{\mu}$ are connected by the thermodynamic relation \eq{thermodynamics} 
and $\chi_{\mu}$ is always larger than $\chi_n$. In other words, $\chi_{\mu}$ is more susceptible, 
which explains the reason why the jump of $\chi_{\mu}$ tends to be pronounced more than $\chi_{n}$.

\bibliography{main} 

\begin{thebibliography}{40}
\bibitem[{mis()}]{misc-exponentTn}
\bibinfo{note}{The singularity of $\frac{\partial \mu}{\partial h}$ at $T_N$ is
  the same as that of $\frac{\partial \Delta_0}{\partial h}$ at $T_c$, but
  their origin is different. It is easy to show $\frac{\partial \mu}{\partial
  h} \propto (T_N-T)^{-\gamma} \times (T_N-T)^{\beta} =
  (T_N-T)^{-(\gamma-\beta)}$ for $T<T_N$ in the vicinity of $T_N$. In
  mean-field theory $\gamma-\beta = 1-1/2=1/2$. See the statement below
  \eq{chin0-} for the singularity of $\frac{\partial \Delta_0}{\partial h}$ at
  $T_c$.}
\end{thebibliography}

\end{document}